\documentclass[aps,prx,superscriptaddress,nofootinbib]{revtex4-1}

\usepackage{mathrsfs}
\usepackage{amsfonts}
\usepackage{amssymb}
\usepackage{amsmath}
\usepackage{graphicx}
\usepackage[usenames,dvipsnames]{color}
\usepackage[colorlinks=true,citecolor=blue,linkcolor=magenta]{hyperref}
\usepackage{ulem}
\usepackage{lmodern}

\newcommand{\Tr}{\mathrm{Tr}}

\newcommand{\abs}[1]{| #1 |}

\newcommand{\ket}[1]{\vert{ #1 }\rangle}

\newcommand{\ketbra}[2]{\vert #1 \rangle \langle #2 \vert}

\begin{document}

%\preprint{APS/123-QED}

\title{A double quantum dot memristor}
\author{Ying Li}
\affiliation{Department of Materials, University of Oxford, 16 Parks Road, Oxford OX1 3PH, UK}
\author{Gregory W. Holloway}
\affiliation{Department of Physics, University of Waterloo, Waterloo, Ontario N2L 3G1, Canada}
\affiliation{Institute for Quantum Computing, University of Waterloo, Waterloo, Ontario N2L 3G1, Canada}
\author{Simon C. Benjamin}
\author{G. Andrew D. Briggs}
\affiliation{Department of Materials, University of Oxford, 16 Parks Road, Oxford OX1 3PH, UK}
\author{Jonathan Baugh}
\affiliation{Institute for Quantum Computing, University of Waterloo, Waterloo, Ontario N2L 3G1, Canada}
\affiliation{Department of Chemistry, University of Waterloo, Waterloo, Ontario N2L 3G1, Canada}
\author{Jan A. Mol}
\email{To whom correspondence should be addressed; E-mail:  jan.mol@materials.ox.ac.uk}
\affiliation{Department of Materials, University of Oxford, 16 Parks Road, Oxford OX1 3PH, UK}

\date{\today}
             
\begin{abstract}
Memristive systems are generalisations of memristors, which are resistors with memory. In this paper, we present a quantum description of quantum-dot memristive systems. Using this model we propose and experimentally demonstrate a simple and practical scheme for realising memristive systems with quantum dots. The approach harnesses a phenomenon that is commonly seen as a bane of nanoelectronics, i.e. switching of a trapped charge in the vicinity of the device. We show that quantum-dot memristive systems have hysteresis current-voltage characteristics and quantum jump induced stochastic behaviour. While our experiment requires low temperatures, the same setup could in principle be realised with a suitable single-molecule transistor and operated at or near room temperature.
\end{abstract} 

\pacs{Valid PACS appear here}

\maketitle

\section{Introduction}

Resistors, capacitors and inductors are the three fundamental circuit elements that are familiar from high school level electronics. Normally, their properties are independent of their history, i.e. they do not exhibit memory effects. In certain complex materials or nanoscale devices, however, memory effects are possible and properties like resistance depend on the states through which the system has evolved \cite{Pershin2011}. A fourth type of two-terminal circuit element characterized by a relationship between charge and flux linkage was described by Chua in 1971 ~\cite{Chua1971}. Predicting that conductance would depend on the history of current flow through it, it was dubbed {\it memristor} for memory and resistor. In 2008, it was shown that memristive behaviour existed in previously observed resistance switching behaviour when ``electronic and ionic transport are coupled", especially in nanoscale films~\cite{2008HP_Nature}. While this was considered the first useful example of a memristor, infact, memristive behaviour characterized by hysteresis under bipolar periodic driving was observed much earlier. It was even argued by Prodromakis et al that memristive behaviour has been observed as early as 200 years ago, as it is characteristic of systems that support electric discharge ~\cite{Prodromakis2012}. The term memristor is now understood in more general terms than first envisaged by Chua, i.e. any two-terminal non-volatile memory devices based on resistance switching, blurring the line with resistive random-access memory RRAM~\cite{Chua1976, Chua2011}.

Subsequent research has realised memristive behaviour in a wide variety of systems, and has identified applications where the memory effect may lead to dramatic advantages in simplicity and power versus conventional electronics. These range from ultra dense, low power memory~\cite{Green2007,KimNanoLet2012} envisaged as a successor to today's RAM and flash, through novel stateful logic processes~\cite{Borghetti2010}, and even neuromorphic computing where the memristor is seen as a synthetic analog of a neuronal synapse~\cite{Jo2010}.

As with many other electronic devices, the behaviour of memristive systems is increasingly affected by quantum phenomena as the device size shrinks to ever smaller dimensions. Memristors based on ionic motion can already be scaled down to less than 10 nm~\cite{Yang2013}, and other candidates of memristive systems include even smaller objects, e.g.~molecules~\cite{Green2007}. At the nanoscale, electron transport through devices is strongly influenced by the discrete energy spectrum and interference effects~\cite{NazarovBook}, which may even emerge at room temperature in certain molecules~\cite{Guedon2012}. It is therefore crucial to understand the non-classical behaviour that emerges in nanoscale memristive systems. A model for a quantum memristor was recently proposed and studied as a potential building block for quantum simulation \cite{Pfeiffer2016}.

In this paper, we present a general quantum-mechanical model of a memristive system and apply it to a specific scheme for a memristive device using two capacitively-coupled quantum dots. Capacitive links between two quantum dots~\cite{Zajac2015} or between a quantum dot and random~\cite{Villis2014} or intentionally placed impurities~\cite{Morello2010} are often used for the read-out of charge- and spin-states in semiconductor nanodevices~\cite{Morello2010}. The effect we now wish to exploit is often observed in nanoelectronic experiments, i.e. the device characteristics alter dramatically when a charge state in its immediate environment changes. This is frequently a frustration in experimental work, but we now propose that it may be usefully harnessed as the basis of memristive function. Room temperature operation of these devices could potentially be achieved in nano-scale semiconductor devices~\cite{Lavieville2015} or by using single-molecule transistors~\cite{Miyamachi2012,Mol2015}. While hysteresis effect have previously been studied in single-electron devices~\cite{Yang2014,Maier2015} and superconducting devices~\cite{Pfeiffer2016,Shevchenko2016,Salmilehto2017}, the use of a double quantum dot system as a memristive device has not been explored to date.

In quantum mechanics, a memristive device can be considered as an open quantum system: the internal state of the device interacts with the environment, typically the electrodes~\cite{Breuer2002, Sun1999}, such that the internal state affects the electron transport (the resistance), and the evolution of the internal state depends on external signals (current and voltage). An example of such an open quantum system is a pair of capacitively-coupled quantum dots, where only one dot participates in the electron transport while the other controls the transport via the capacitive coupling. The environment (electrodes) is configured such that the evolution of the control dot depends on the applied voltage. We will show that such a double-dot device has the characteristic property of memristive systems, i.e.~current-voltage hysteresis curves~\cite{Chua1976}. Following the theoretical description of this open quantum system, we demonstrate an experimental realization of the hysteresis in a capacitively coupled pair of silicon metal-oxide-semiconductor (MOS) quantum dots. The current in quantum memristive systems shows stochastic behaviour and may not converge to a periodic curve under periodic driving due to quantum jumps between different values of the quantised conductance. Stochastic behaviour has also been observed in some classical memristive systems \cite{Gaba2013}, but in the quantum-dot memristor examined here, it is due to single-electron tunneling.

%\section*{Results}

%\subsection*{Quantum description of memristive systems}

\begin{figure}[tbp]
\centering
\includegraphics[width=0.55\linewidth]{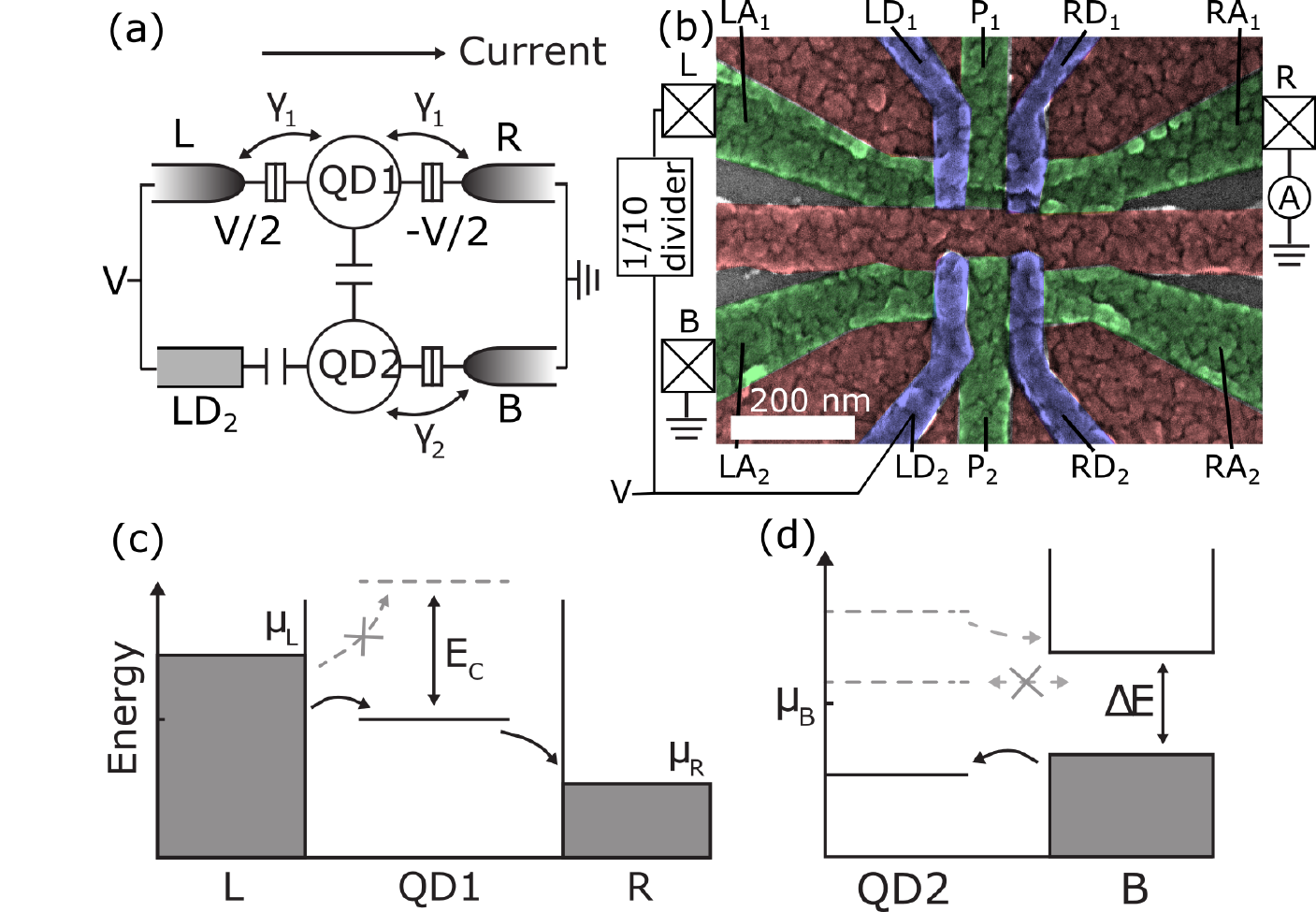}
\caption{
\textbf{Quantum-dot memristive system.} (a) Circuit schematic of the memristive system with two terminals: reservoir-L and reservoir-R. When the voltage $V$ is applied on the device, electrochemical potentials in these two reservoirs are respectively $\mu_\text{L} = \abs{e}V/2$ and $\mu_\text{R} = -\abs{e}V/2$. Two terminals are coupled to the quantum dot QD1 via tunnelling with the strength $\gamma_1$, hence current can go through the device via QD1. The quantum dot QD2 is capacitively coupled to QD1 and reservoir-L, and coupled to reservoir-B via tunnelling with the strength $\gamma_2$. The electrochemical potential in reservoir-B is $\mu_\text{B} = 0$. (b) False color SEM image showing the three layers of aluminum gates used to define a capacitively coupled pair of quantum dots in undoped silicon. The center-to-center distance between the two dots is $\sim$ 140 nm. The letters along the top and bottom label the different gates, and the letters along the sides indicate the leads accumulated in the Si. For each dot there are two accumulation gates (LA, RA), two depletion gates (LD, RD), and a plunger gate (P). A dot is formed under each plunger gate where it intersects the two accumulation gates. The lines outside of the image indicate how the device is connected to the rest of the measurement circuit, in order to match the implementation shown in (a). (c) The current is switched on when the energy level of QD1 is between $\mu_\text{L}$ and $\mu_\text{R}$ and off when the energy level of QD1 is above $\max \{\mu_\text{L} , \mu_\text{R} \}$ or below $\min \{\mu_\text{L} , \mu_\text{R} \}$. (d) If there is an energy gap around the Fermi energy in the spectrum of reservoir-B, QD2 can be loaded with an electron only when the energy level is lower than the gap and unloaded only when the energy level is above the gap. 
}
\label{fig:setup}
\end{figure}

First we will give a generic description of a quantum-mechanical memristive system. Any voltage-controlled memristive system is defined by equations~\cite{Chua1976}
\begin{eqnarray}
I &=& G(x,V,t)V \label{eq:MSE1}, \\
\frac{dx}{dt} &=& f(x,V,t) \label{eq:MSE2},
\end{eqnarray}
where the conductance $G$ depends on the voltage $V$ and the state parameter $x$. In such a memristive system, the evolution of the state is controlled by the voltage, and the state equation [Eq.~(\ref{eq:MSE2})] is assumed to have a unique solution for any initial condition.

For a quantum system, the state is described by a reduced density matrix $\rho$. If $\rho$ is the state of the total system including both the system (device) and the environment (e.g.~electrodes coupled to the device), the current going through the device reads
\begin{eqnarray}
I = \Tr ( \hat{I} \rho ),
\end{eqnarray}
where the current operator $\hat{I}$ is an operator of the total system. When the coupling between the system (device) and the environment is weak, the influence of the system on the environment is small, and the state of the total system is approximately a tensor product $\rho \approx \rho_\text{S} \bigotimes \rho_\text{E}$, where $\rho_\text{S}$ ($\rho_\text{E}$) is the state of the system (environment)~\cite{Breuer2002}. Under the weak-coupling condition, the current is approximately
\begin{eqnarray}
I \approx \Tr ( \hat{I}_\text{S} \rho _\text{S} ), \label{eq:QMSE1}
\end{eqnarray}
where $\hat{I}_\text{S} = \Tr _\text{E} ( \hat{I} \openone_\text{S} \bigotimes \rho_\text{E} )$ is an operator of the device depending on the state of the environment, and $\openone_\text{S}$ is the identity operator of the system. Because the state of the environment depends on the voltage, i.e.~electrochemical potentials in electrodes, the operator $\hat{I}_\text{S}$ is voltage-dependent.

For a memristive system, the current is zero whenever the voltage is zero, regardless of the state of the device, which is one of the criteria distinguishing a memristive system from an arbitrary dynamical system~\cite{Chua1976}. Therefore, if $I \vert _{V=0} = 0$ is satisfied, Eq.~(\ref{eq:QMSE1}) is a quantum-mechanical version of Eq.~(\ref{eq:MSE1}), in which the conductance $G = I/V$ depends on both the voltage and the state of the device.

The evolution of a Markov open quantum system is given by the master equation~\cite{Breuer2002}
\begin{eqnarray}
\frac{d\rho _\text{S}}{dt} &=& \mathcal{L}(V) \rho _\text{S} \label{eq:QMSE2},
\end{eqnarray}
which has a unique solution for any initial condition. Here, the generator $\mathcal{L}$ of the semigroup is determined by the state of the environment, i.e.~the voltage $V$. Therefore, when the condition $I \vert _{V=0} = 0$ is satisfied, Eqs.~(\ref{eq:QMSE1})~and~(\ref{eq:QMSE2}) describe a memristive system in quantum mechanics.

In the next section, we will give an example of memristive systems described by Eqs.~(\ref{eq:QMSE1})~and~(\ref{eq:QMSE2}), which is based on quantum dots.

%\subsection*{Quantum-dot memristive system}

\begin{figure}[tbp]
\centering
\includegraphics[width=0.9\linewidth]{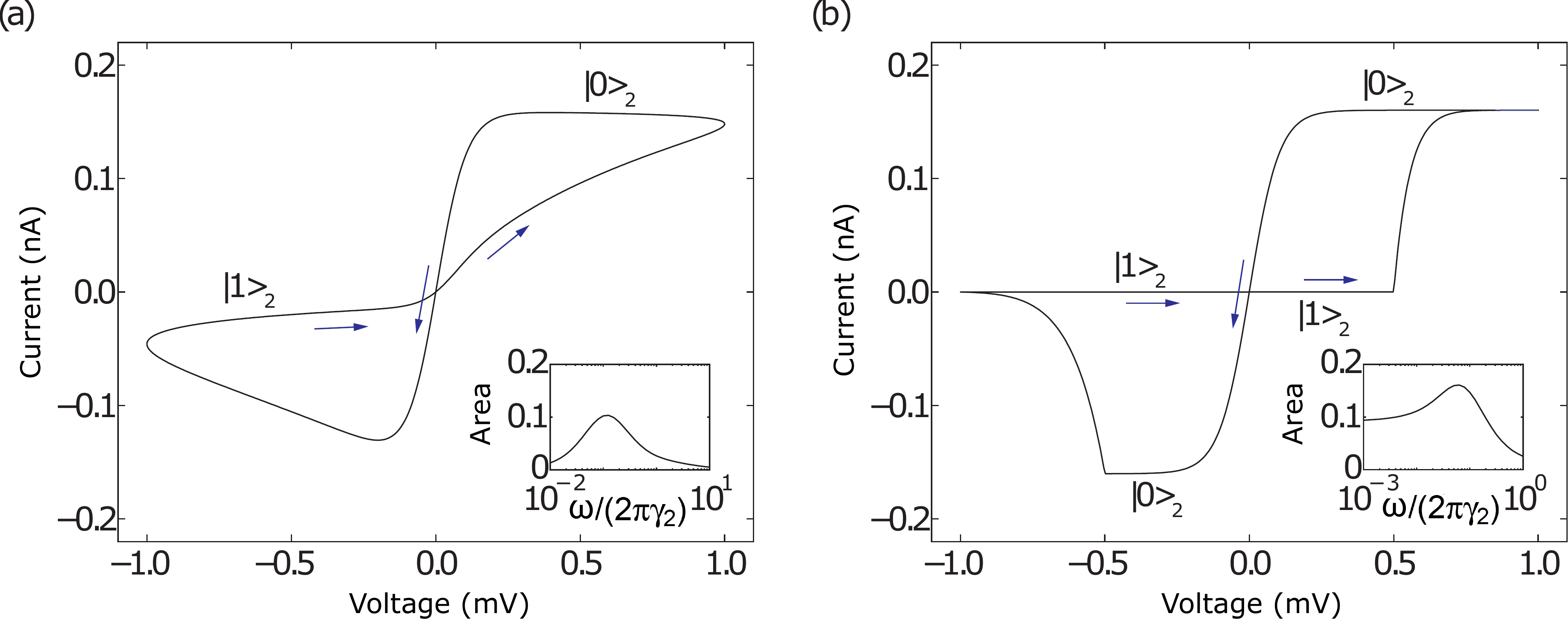}
\caption{
\textbf{Memristive behaviour.} Current-voltage curves of the quantum-dot memristive system driven by a periodic voltage. The voltage is in the form $V(t) = V_\text{A}\cos{\omega t}$, where $V_\text{A} = 1\text{ meV}$ and $\omega = 2\pi \times 1\text{ Hz}$. We have set that the temperature is $T = 0.3\text{ K}$, the capacitive coupling between two quantum dots is $E_\text{C} = 1\text{ meV}$, the tunnelling coupling of QD1 is $\gamma_1 = 2\text{ GHz}$, and the lever arm of reservoir-L gating QD2 is $1$. Average current through QD1 is obtained by numerically solving the master equation. (a) Reservoir-B has a closed energy gap, i.e.~$\Delta E = 0$, and the tunnelling coupling of QD2 is $\tilde{\gamma}_2 = 10\text{ Hz}$. QD2 is likely empty ($\ket{0}_2$) when the voltage is decreasing but still positive, occupied ($\ket{1}_2$) when the voltage is increasing but still negative and in a mixed state in other cases. (b) Reservoir-B has an energy gap $\Delta E = 0.5\text{ meV}$ centred at the energy $0$, and the tunnelling coupling of QD2 is $\tilde{\gamma}_2 = 100\text{ Hz}$. When the energy level of QD2 is within the energy gap, i.e.~$V \in [-0.5\text{ meV},\text{ }0.5\text{ meV}]$, QD2 is likely empty when the voltage is decreasing and occupied when the voltage is increasing. Insets show the area of the hysteresis loops in units of mV $\times$ nA as a function of the ratio of driving frequency to QD2 tunneling rate. When $\Delta{E} = 0$ the area approaches zero for driving frequencies that are significantly larger or smaller than the tunneling rate. In contrast, for finite $\Delta{E}$ the area reaches a fixed value when $\omega/(2\pi\tilde{\gamma}_2) \ll 1$.
}
\label{fig:hysteresis}
\end{figure}

We consider a system with two quantum dots as shown in Fig.~\ref{fig:setup}(a). Quantum dots have discrete energy spectrum due to the confinement in all three directions. Coulomb repulsion between electrons in a dot leads to an increasing energy cost for adding more electrons to the dot. When the energy space between discrete states and the Coulomb repulsion energy are large compared to any thermal fluctuations, each quantum dot can be treated as a single-level system, i.e.~each dot has only two states: empty or occupied with one electron.

In our device, the two terminals of the memristive system are electron reservoirs L and R [see Fig.~\ref{fig:setup}(a)], which are formed beneath the aluminum accumulation gates LA$_1$ and LA$_2$ [see Fig.~\ref{fig:setup}(b)]. These terminals are coupled to the quantum dot QD1 with tunnelling strengths that are determined by the depletion gates LD$_1$ and RD$_1$. First, we will consider the case where we have an equal tunneling strength $\gamma_1$ to terminals L and R. The input of the memristive system is the voltage $V$ applied on these two terminals, and the output is the current $I$ going through QD1. We set the electrochemical potential in the reservoir-L (reservoir-R) $\mu_\text{L} = \abs{e}V/2$ ($\mu_\text{R} = -\abs{e}V/2$). When the energy level of QD1, which is controlled by the plunger gate P$_1$, is within the window defined by $\mu_\text{L}$ and $\mu_\text{R}$ [see Fig.~\ref{fig:setup}(c)], electrons can flow through QD1 via tunnelling couplings to the terminals; when the energy level of QD1 is outside of the window, the current decreases exponentially with the energy due to the Fermi-Dirac distribution of electrons in the terminals. Therefore, the current can be switched on and off by changing the energy level of QD1. When QD1 is in the steady state, the current is~\cite{Sun1999}
\begin{eqnarray}
I = \abs{e}\gamma_1 [ f(E_1-\mu_\text{L}) - f(E_1-\mu_\text{R}) ]/2,
\end{eqnarray}
where $E_1$ is the energy of the QD1 level, $f(E) = [ 1 + \exp(E/k_\text{B}T) ]^{-1}$, and $T$ is the temperature.

The second quantum dot QD2 is capacitively coupled to QD1. When the quantum dot QD2 is empty, we set the energy of the QD1 level $E_1 = 0$; when QD2 is occupied, the energy of the QD1 level is raised to $E_1 = E_\text{C}$ due to the capacitive coupling. Therefore, if the capacitive coupling is strong enough, i.e.~$E_\text{C} > \max \{ \mu_\text{L},\mu_\text{R} \}$, the current is switched on and off depending on the occupation in QD2. When QD1 is in the steady state, conductances are
\begin{eqnarray}
G_\text{ON} = \abs{e}\gamma_1 [ f(-\mu_\text{L}) - f(-\mu_\text{R}) ]/(2V),
\label{eq:Gon}
\end{eqnarray}
and
\begin{eqnarray}
G_\text{OFF} = \abs{e}\gamma_1 [ f(E_\text{C}-\mu_\text{L}) - f(E_\text{C}-\mu_\text{R}) ]/(2V),
\label{eq:Goff}
\end{eqnarray}
respectively, for an empty and occupied QD2.

The relaxation time of QD1 is $\sim \gamma_1^{-1}$. We will focus on the case where the voltage is switched with a frequency much lower than $\gamma_1$. In this case, the average conductance is approximately given by $G = (1-\bar{n}_2) G_\text{ON} + \bar{n}_2 G_\text{OFF}$, where $\bar{n}_2$ is the average occupation of QD2. The conductance is always finite, i.e.~the condition $I \vert _{V=0} = 0$ is satisfied.

In order to have a voltage-dependent evolution of the state, QD2 is capacitively coupled to the left terminal by galvanically connecting the electron reservoir beneath accumulation gates LA$_1$ and one of the gates of QD2. Hence, the energy level of QD2 is controlled by the voltage. We assume that when QD1 is empty, the energy of the QD2 level is $E_2 = \mu_\text{L}$. QD2 is also coupled to the reservoir-B, which is beneath accumulation gate LA$_2$, with the tunnelling strength $\gamma_2$ determined by depletion gate LD$_2$ [see Fig.~\ref{fig:setup}]. The reservoir-B is a bath of electrons, which allows the state of QD2 to be switched between empty and occupied. Due to the tunnelling coupling, QD2 is emptied with the rate
\begin{eqnarray}
\gamma_\text{out} = \gamma_2(E_2) [1-f(E_2-\mu_\text{B})], \label{eq:gammaOUT}
\end{eqnarray}
and occupied with the rate
\begin{eqnarray}
\gamma_\text{in} = \gamma_2(E_2) f(E_2-\mu_\text{B}) \label{eq:gammaIN}
\end{eqnarray}
where $\mu_\text{B} = 0$ is the electrochemical potential in the reservoir-B. Here, we have assumed that the tunnelling coupling strength $\gamma_2$ can depend on the energy of the QD2 level, e.g.~when there is an energy gap in the spectrum of the reservoir-B, the tunnelling vanishes in the energy gap. Therefore, the evolution of QD2 is controlled by the voltage, through either the tunnelling rate or the position of the QD2 energy level relative to $\mu_B$.

%\subsection*{Hysteresis {\it I-V} curves}

We consider the case where the input signal is an oscillatory voltage applied on electron reservoirs L and R of our memristive device. In the limit of low input frequency, the device is always in the steady state, and the occupation of QD2 usually depends only on the instantaneous voltage. Therefore, $I$ is a single-valued function of $V$ independent of the sweep direction. In the limit of high input frequency, the state of the device is only determined by the initial state, yielding an {\it I-V} curve with no hysteresis. This is because the sweep duration is much shorter than the tunnelling time, preventing a tunnelling event from occurring. However, for a moderate input frequency, the state cannot achieve the steady state of the instantaneous voltage before the voltage changes and thence evolves periodically, which results in a double-valued {\it I-V} function, i.e.~the {\it I-V} curve is a Lissajous figure with two loops joined at $(I=0,V=0)$.

\begin{figure}[!t]
\centering
\includegraphics[width=0.55\linewidth]{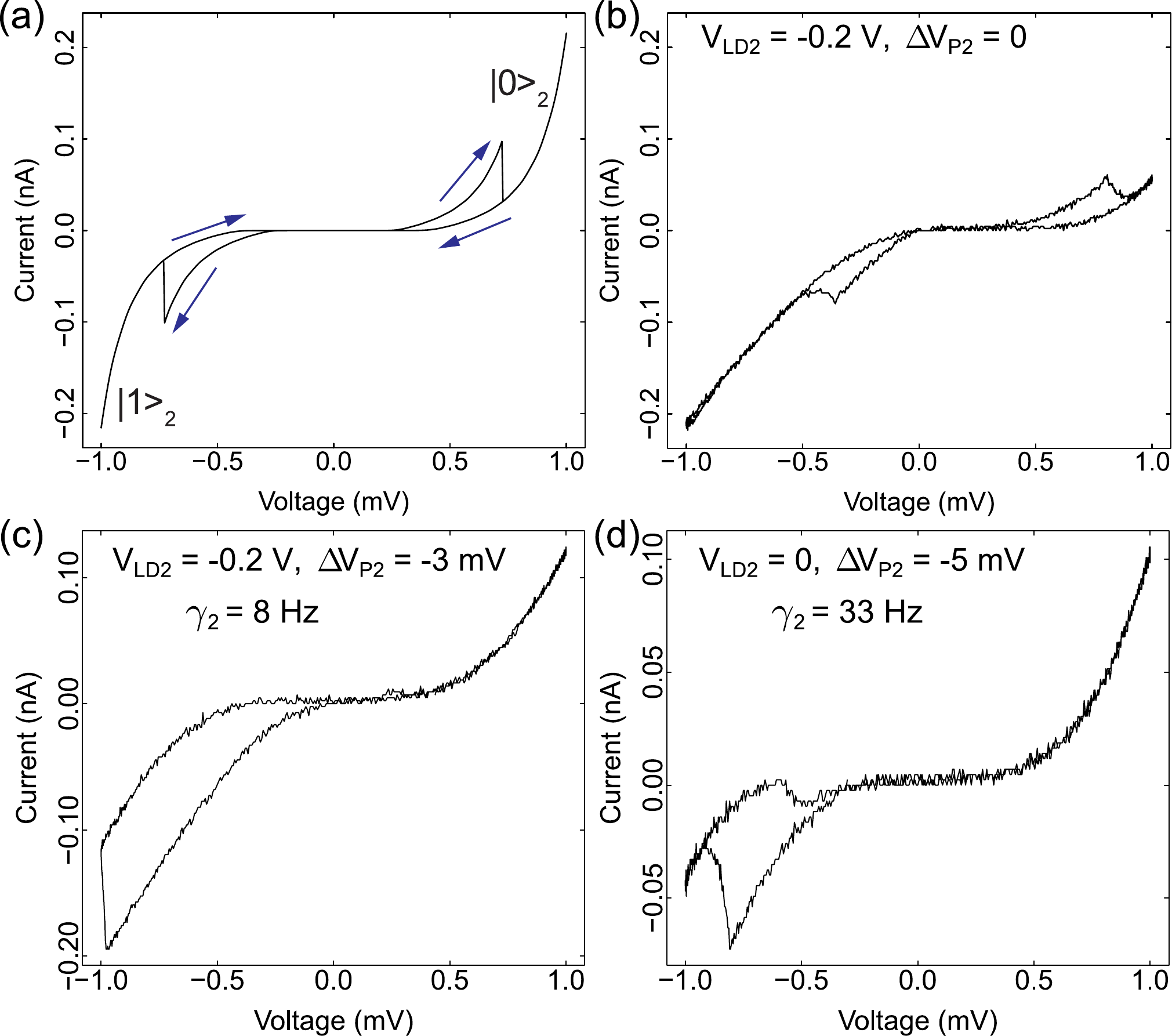}
\caption{
\textbf{Hysteresis traces.} Simulated and experimental {\it I-V} curves showing current hysteresis in the Si double dot system shown in Fig.~\ref{fig:setup}(b). (a) Simulated {\it I-V} curve of QD1 for one sinusoidal voltage cycle with $V_A = 1$ mV. The suppression of current near zero bias is due to a weak, unintentional potential barrier in series with QD1. The blue arrows relate the curves to the voltage sweep direction. $\ket{0}_2$ and $\ket{1}_2$ indicate the charge state of QD2. (b-d) Measured current through QD1 for one voltage cycle with $\omega = 2\pi \times 1$ Hz. (b) Two-looped hysteresis showing similar behavior as in (a). (c) The position of the hysteresis is shifted to negative bias by lowering V$_{P2}$ which changes the energy level $E_2$ in QD2. $\Delta{V}_{P2}$ denotes the difference in voltage applied to $V_{P2}$, relative to when the hysteresis is symmetric about zero bias. (d) Raising the DC voltage on LD$_2$ lowers the tunnel barrier and increases $\gamma_2$, causing the hysteresis width to narrow. The estimated tunnel rates $\gamma_2$ for (c) and (d) are 8 Hz and 33 Hz, respectively.
}
\label{fig:results}
\end{figure}

In the quantum-dot memristive system, QD2 provides the memory of the device, and the evolution of QD2 is determined by reservoir-B. If reservoir-B has an energy gap in the spectrum as shown in Fig.~\ref{fig:setup}(d), QD2 can be loaded with an electron only when the energy level is below the energy gap and unloaded only when the energy level is above the energy gap. Physical realization of such a gap could be achieved in superconductors or systems with irregular density of states due to disorder and confinement. Driving the energy level of QD2 with a periodic voltage, the conductance is $G_\text{ON}$ if the energy level is moving downward and $G_\text{OFF}$ if the energy level is moving upward. At a finite temperature, energy-dependent unloading and loading rates of QD2 [Eqs.~(\ref{eq:gammaOUT})~and~(\ref{eq:gammaIN})] can be caused by the Fermi-Dirac distribution of electrons in reservoir-B (with the theoretical 'best case' at zero temperature). This mechanism results in a double-valued {\it I-V} function as shown in Fig.~\ref{fig:hysteresis}(a). In the presence of an energy gap $\Delta E$, the hysteresis becomes more marked. Fig.~\ref{fig:hysteresis}(b) show how an energy gap as small as $1\text{ meV}$ qualitatively changes the shape of the hysteresis curve, with the tunnelling coupling between QD2 and reservoir-B $\gamma_2(E_2) = \tilde{\gamma}_2[\theta(-\Delta E/2 - E_2) + \theta(E_2 - \Delta E/2)]$. Larger energy gaps further extend the range of voltage over which hysteresis occurs. Here, $\Delta E$ is the energy gap, and $\theta(E)$ is the Heaviside step function.

Without the energy gap, the relaxation time of QD2 needs to be comparable to the period of the driving signal, otherwise the hysteresis disappears [see inset of Fig.~\ref{fig:hysteresis}(a)]. In the presence of an energy gap this requirement is less stringent, since the relaxation of QD2 is suspended when the energy level of QD2 is in the gap. Therefore the hysteresis feature persists when $\omega/(2\pi\tilde{\gamma}_2) \ll 1$,  with a shape independent of the value of $\tilde{\gamma}_2$. [see inset of Fig.~\ref{fig:hysteresis}(b)]. In both cases, if $\omega/(2\pi\tilde{\gamma}_2) \gg 1$ the area of the hysteresis loops goes to zero. This is due to the evolution of QD2 being much slower than the driving signal, such that the state of QD2 is frozen, and the current response stays the same regardless of the voltage sweep direction. Additionally, the hysteresis dims when the temperature increases. The hysteresis relies on the energy-dependent couplings $\gamma_\text{out}$ and $\gamma_\text{in}$. Without the energy gap, the energy-dependent coupling is due to the Fermi-Dirac distribution, which becomes flattened when the temperature increases. In the presence of an energy gap, $\gamma_2(E_2)$ always varies significantly at edges of the gap. However, if the temperature is higher than the gap, $\gamma_\text{out}$ and $\gamma_\text{in}$ are similar on both sides of the gap, i.e.~the state of QD2 does not evolve significantly when the energy level is swept from one side of the gap to another.

We have discussed the hysteresis based on the coupling to reservoir-B. The hysteresis can also be caused by a finite tunnelling coupling between two quantum dots. Due to the tunnelling coupling, the orbit of QD2 is extended and overlaps with QD1, such that QD2 could also be coupled to the two terminals (reservoirs L and R). If the interdot tunneling rate is larger than $\gamma_2$, tunneling to QD1 provides an alternate relaxation path for QD2, and the two terminals will replace the role of reservoir-B. However, since $\gamma_1$ is generally faster than $\gamma_2$, QD2 will achieve the instantaneous equilibrium state very quickly, suppressing the hysteresis effect. Therefore, it is necessary to keep the interdot tunneling rate low, such that the relaxation rate of QD2 is comparable with the driving frequency, and hysteresis can be observed. Experimentally, this is achieved by grounding the red gate between the two dots in Fig.~\ref{fig:setup}(b), creating a large tunnel barrier which suppresses interdot tunneling. When the relaxation rate of QD2 is comparable with the driving frequency, we expect to see the hysteresis.

%\subsection*{Experimental results}

We implemented the above scheme (for no energy gap in reservoir-B) using the parallel-configuration double quantum dot device shown in Fig.~\ref{fig:setup}(b). We use the same nomenclature of QD1 and QD2 to indicate the two quantum dots of the experimental device. To form QD2, we lower both depletion gates LD$_2$ and RD$_2$ to $\sim 0.1$ V, forming high tunnel barriers. Since direct transport through QD2 cannot be measured with such high barriers, the hysteresis in the QD1 current is used to estimate the tunnel rate $\sim 10$ Hz (see below). In this gate range, we observe that LD$_2$ is able to modulate the tunnel rate, whereas RD$_2$ is not, from which we conclude that QD2 is formed closer to the left side and tunnelling from the right lead is absent (hence reservoir-B is indicated to the left of QD2 in Fig.~\ref{fig:setup}(b). To control the QD2 level E$_2$ during the AC voltage sweep, we also use gate LD$_2$, since it is strongly coupled to the dot. While LD$_2$ is used both for the AC sweep and for tuning the tunnel rate $\gamma_2$, we note that the tunnel rate is practically insensitive to LD$_2$ over the AC sweep range of millivolts. It is important to note that in this experiment a voltage divider was used to increase the magnitude of the AC voltage on LD$_2$ by a factor of 10 compared to reservoir-L. That is, a 10 mV AC signal is applied to LD$_2$ but is reduced to $V_A = 1$ mV on reservoir-L by the divider. This is because the lever arm between LD$_2$ and QD2 is of the order 0.2, but the hysteresis effect is more evident when the potential felt by both dots is comparable. The results are summarized in Fig.~\ref{fig:results}, where we show the {\it I-V} responses for a single voltage cycle (in contrast to the average current presented in Fig.~\ref{fig:hysteresis} that would correspond to an average over many cycles). For all panels of Fig.~\ref{fig:results}, a sweep rate of $\omega = 2\pi \times 1$ Hz and sweep amplitude $V_A = 1$ mV are used, and the current through QD1 is displayed. Fig.~\ref{fig:results}(b) shows hysteresis that is symmetric about $V=0.2$, indicating that energy $E_2$ of QD2 is aligned with $\mu_B$ at 0.2 $V$. Fig.~\ref{fig:results}(a) is a simulated trace using Eqs.~(\ref{eq:Gon})~and ~(\ref{eq:Goff}) with instantaneous switching of the charge state of QD2 chosen to match the data in panel (b). For a single cycle, sharp jumps in current occur when the charge state of QD2 changes. While these jumps are perfectly sharp in simulation (Fig.~\ref{fig:results}(a)), they are slightly rounded in the experiment [Fig.~\ref{fig:results}(b-d)] due to the limited bandwidth of the RC filtering employed in our experimental circuit.

Tuning the DC voltages applied to QD2 varies the hysteresis shape in two distinct ways. The plunger gate P$_2$ controls the value of $E_2$. This changes the position of the dot level relative to $\mu_B$, so that a different value of $\mu_L$ is required to change the charge state of QD2. The position of the hysteresis loop is thus shifted along the {\it I-V} curve, as shown in Fig.~\ref{fig:results}(c). This confirms that the observed jumps in the current of QD1 are due to charge transitions on QD2. If QD2 is detuned far enough, the voltage $V$ will no longer be sufficient to cause the dot level to cross $\mu_B$, and no hysteresis will be observed. 

Current is suppressed at low bias due to a weak unintentional tunnel barrier in series with QD1. This device imperfection is included in the model by considering the dot and tunnel barrier as two voltage dependent resistors in series. The {\it I-V} curve is calculated from the total resistance, which changes as a function of $V$ depending on how the voltage is dropped across the two resistors. The resistance of the dot $R_d(V_d)$ is calculated from Eqs.~(\ref{eq:Gon})~and ~(\ref{eq:Goff}) using $R = 1/G$. For the tunnel barrier, resistance $R_b(V_b)$ is calculated for transmission through a square barrier with a height of 0.3 meV, and a width of 40 nm (best fit parameters). $V_d$ and $V_b$, the voltages dropped across the dot and barrier respectively, are related to the bias voltage by: $V = V_d+V_b$. Comparison of Fig.~\ref{fig:results}(a) and Fig.~\ref{fig:results}(b) shows that this model reproduces the experimental data well. If we remove the effect of the tunnel barrier resistance and plot the I-V response of the double dot system only, a curve similar to Fig.~\ref{fig:hysteresis}(a) is obtained, but with stochastic jumps rather than an average current. 

\begin{figure}[!tbp]
\centering
\includegraphics[width=0.45\linewidth]{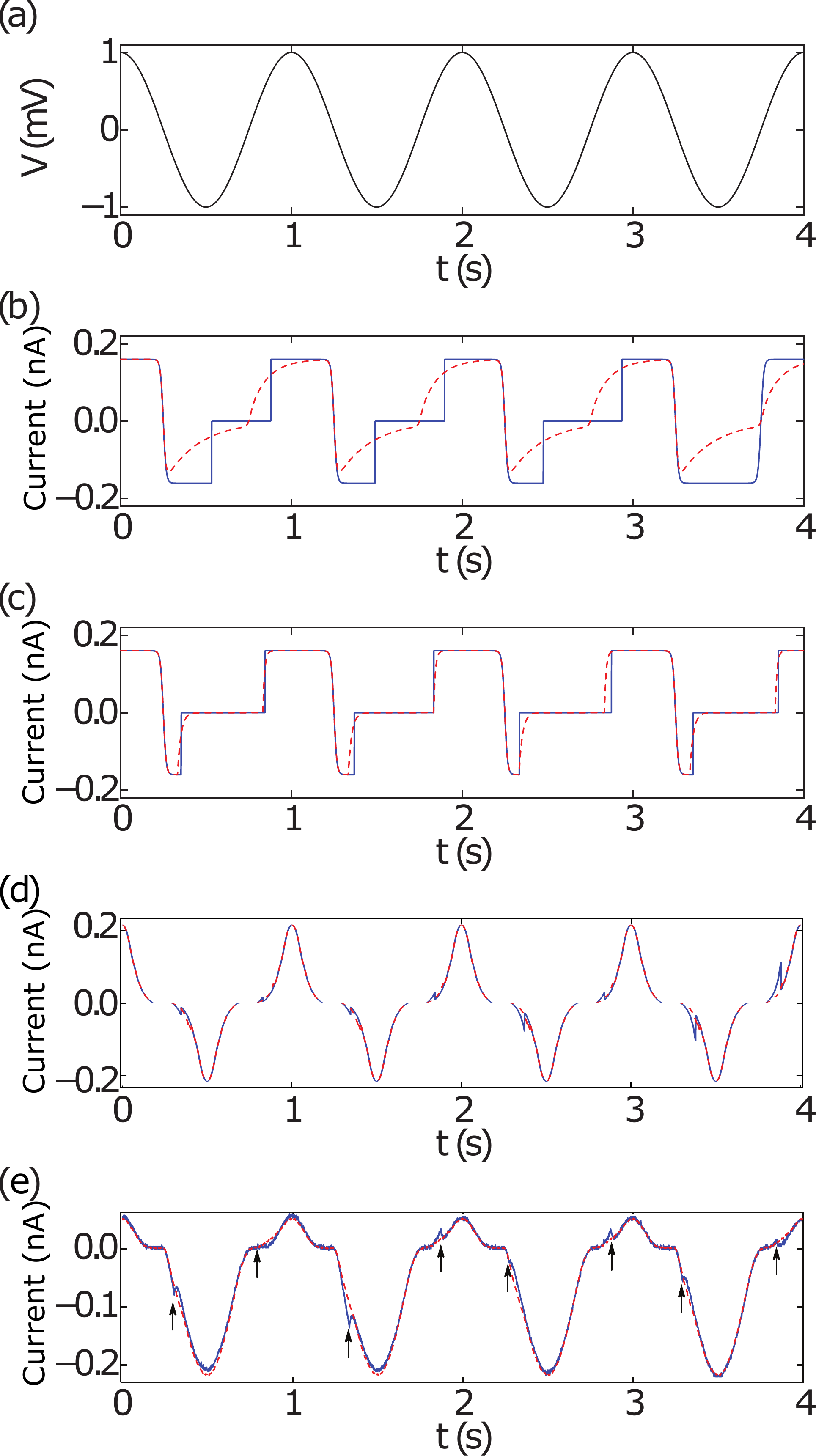}
\caption{
\textbf{Quantum jumps} Quantum trajectory simulations of the instantaneous current in the quantum-dot memristive system. Parameters are the same as in Fig.~\ref{fig:hysteresis} (a). (a) Voltage applied on two terminals. The energy gap in reservoir-B is closed in (b) and open in (c), i.e.~parameters of simulated systems are the same as in Fig.~\ref{fig:hysteresis}~(a)~and~(b), respectively. The instantaneous current (blue solid curve) is different from the current averaged over many rounds (red dashed curve) due to quantum jumps, which correspond to sudden changes of the instantaneous current. (d) Simulated instantaneous and average current for the memristive system with the weak additional tunnel barrier included. Parameters of the simulation are the same as those used in Fig.~\ref{fig:results}(a). (e) Experimental data. Blue curve is instantaneous current and the red dashed curve is the current averaged over 10 sweeps. Black arrows indicate the position of the abrupt current jumps. The jump positions vary stochastically from sweep to sweep, consistent with the expected behavior of quantum jumps
}
\label{fig:jump}
\end{figure}

Since reservoir-B does not have an energy gap in this device, we require that the tunnelling rate $\gamma_2$ must be in the correct frequency range in order to observe the memristive hysteresis effect, i.e. roughly comparable to sweep rate $\omega$. The width of the hysteresis loop is modulated by adjusting $\gamma_2$, using the DC voltage on the depletion gate LD$_2$. This can be seen in Fig.~\ref{fig:results}(c,d), where a more positive voltage in (d) increases the tunnelling rate and the charge transitions occur closer together. If the tunnel rate were further increased the hysteresis would eventually disappear. The rate $\gamma_2$ can be estimated for a given {\it I-V} curve as the ratio of the average voltage sweep rate to the distance between the two charge transitions: $\gamma_2 = 2V_a\omega/(\pi\Delta{V}_{\rm hyst})$, where $\Delta{V}_{\rm hyst}$ is the hysteresis width. For the curves in Fig.~\ref{fig:results}(c,d), tunnelling rates of 8 Hz and 33 Hz are estimated, respectively.

\begin{figure}[tbp]
\centering
\includegraphics[width=0.9\linewidth]{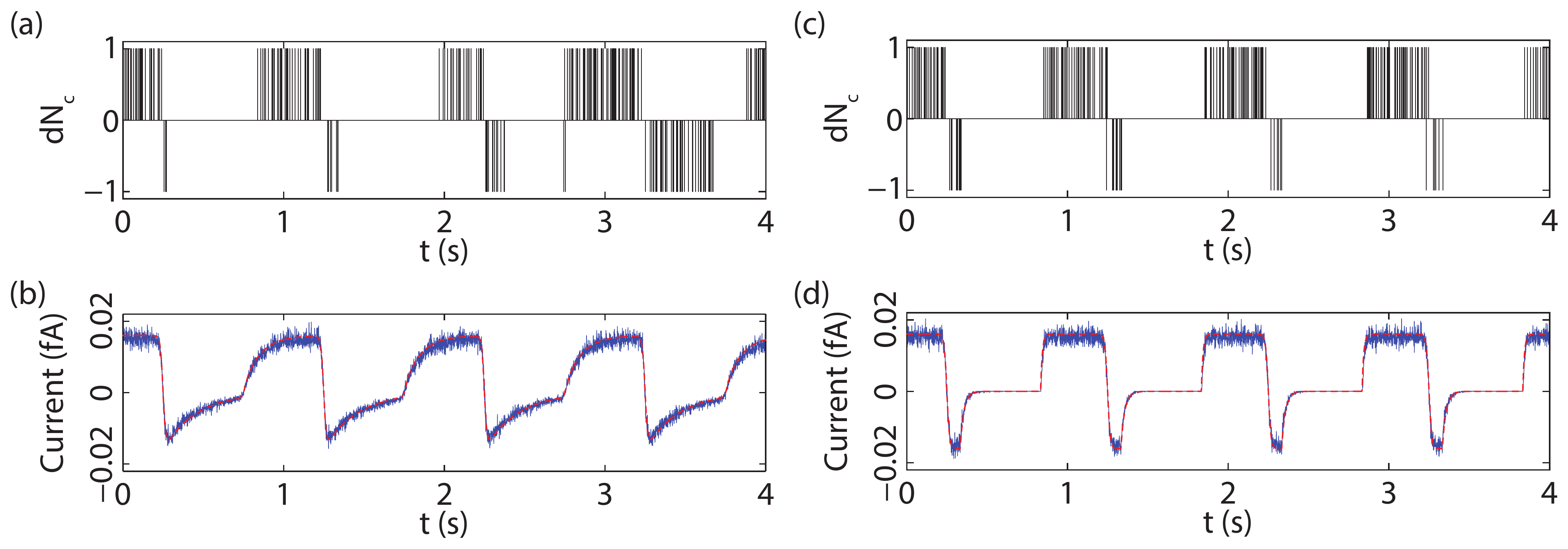}
\caption{
\textbf{Electron tunneling} Quantum trajectory simulations of electron tunnelling in the quantum-dot memristive system. The tunnel coupling of QD1 is $\gamma_1 = 200\text{ Hz}$, whereas other parameters are the same as in Fig.~\ref{fig:hysteresis}~(a)~and~(b). The energy gap in reservoir-B is closed in (a) and (b) and open in (c) and (d). (a) and (c) show electron tunnelling events from QD1 to reservoir-R in a single trial. $N_{\rm c}$ is the total number of electrons tunnelling from QD1 to reservoir-R. Each vertical line labels a tunnelling event: when $dN_{\rm c} = 1$ the electron tunnels from QD1 to reservoir-R; when $dN_{\rm c} = -1$ the electron tunnels from reservoir-R to QD1. In figure (a), there are two tunnelling events in the wrong direction (near $t=2.8$ s) due to thermal fluctuation, and we expect more similar events at higher temperatures. (b) and (d) show the current averaged over $1000$ trials (blue solid curve)~\cite{footnote}, which converges to the red curve when the number of trials goes to infinity.}
\label{fig:trajectory}
\end{figure}

%\subsection*{Quantum jumps}

{\it I-V} curves in Fig.~\ref{fig:hysteresis} are ensemble averages of many periods, in which the average current varies continuously. Because the instantaneous conductance is quantised and only has two possible values $G_\text{ON}$ and $G_\text{OFF}$, the instantaneous current could be different from the average value and change stochastically due to quantum jumps between the empty and occupied states of QD2. In Fig.~\ref{fig:jump}(b,c), we numerically simulate the instantaneous current using the quantum trajectory method~\cite{Goan2001}. If reservoir-B has an energy gap, as in Fig.~\ref{fig:jump}(c), quantum jumps only occur when the energy level of QD2 is around the edge of the gap. In this case, the state of QD2 is always determined [also see Fig.~\ref{fig:hysteresis}(b)], i.e.~the instantaneous current and the average current are nearly the same. However, if the energy gap of reservoir-B is closed, as shown in Fig.~\ref{fig:jump}(b), the occupation of QD2 is probabilistic about half of the time [see Fig.~\ref{fig:hysteresis}(a)], hence the difference between the instantaneous current and the average current is more apparent. Because of the stochastic jumps, we can anticipate that the instantaneous current would never converge to a periodic curve. Experimental data in Fig.~\ref{fig:jump}(d) confirms that the position of the charge transitions is not the same in each voltage sweep, and the instantaneous and average currents are not equal. The occurrence of quantum jumps is an intrinsic feature of quantum-mechanical memristive devices. In contrast to classical stochastic memristors, here we have control over the tunnel rates and can alter the statistical distribution of tunnelling events.

When the tunnelling coupling of QD1 is weak, the shot noise modifies the {\it I-V} curves. In Fig.~\ref{fig:trajectory}, we numerically simulate the tunnelling of electrons through QD1~\cite{Goan2001}. Even in a single trial, the asymmetry of the electron-count distribution in one period reflects the double-valued {\it I-V} characteristic of the memristive system. In a real circuit the shot-noise as illustrated in Figs.~\ref{fig:trajectory}(a)~and~(c) is unlikely to be observed; instead of stochastic jumps, there is a smooth stochastic current due to the finite response time of the circuit.

%\section*{Discussion}

We have presented a model of quantum memristive systems based on a quantum master equation approach. Inspired by this model, we proposed a specific scheme of memristive systems built with quantum dots, and demonstrated the effect experimentally in a capacitively-coupled, parallel double quantum dot device. Numerical results showed that with an energy gap in the spectrum of the bath reservoir, quantum-dot memristive systems can be used as non-volatile memories. Our results demonstrate single-electron memristive behavior at cryogenic temperatures, however, systems with charging energies of several hundreds of meV\cite{Lavieville2015,Mol2015} should enable room temperature operation.

%Our results demonstrate that, in addition to the hysteresis {\it I-V} characteristics of memristive systems, these quantum-dot devices behave non-classically. They provide an additional resource in the catalogue of circuit components for quantum technologies.

\section*{Methods}

\subsection*{Device fabrication}

Devices were fabricated using undoped silicon with a thermal oxide of 300 nm, which was etched down to a thickness of 12 nm in the area where the dots are patterned. Ion implantation of phosphorus with an energy of 12 keV was performed through a photomask to create degenerately doped source/drain contacts. Following implantation, the resist mask was stripped, and the chip was annealed at 950 $^\circ$ C in N$_2$ to remove damage from the implantation process, and to activate the dopants. To prevent shorts between these ion implanted regions and subsequent metal gates, a 6 nm layer of Al$_2$O$_3$ was deposited by atomic layer deposition. Gates to define the quantum dots were then fabricated on the substrate using three layers of Al gates. Each layer was patterned using electron beam lithography with a PMMA mask, and Al deposited using an electron beam evaporator with a deposition rate of 1 \AA/s. Following liftoff of the excess metal, the devices were exposed to a 25 W O$_2$ plasma at 150 $^\circ$C for two minutes to form an oxide layer at the metal surface. This oxide provides electrical insulation between gate layers, and we find this process results in gate-gate breakdown voltage greater than 4 V. The first gate layer, shown in red in figure 1a, is a screening layer which controls where subsequent gates will form an accumulation layer at the Si/SiO$_2$ interface. The screening gates define two horizontal channels with a separation of $\sim 140$ nm. Layer two, shown in green, are the accumulation gates used to form the conducting channel between the dots and the source/drain contacts, as well as the plunger gates which control the chemical potentials of the dots. The blue layer is deposited last and is used to form tunnel barriers. After all gate layers are deposited, the chip is annealed in forming gas at 220 $^\circ$C for 1 hour, which is found to be critical for obtaining clean and reproducible device characteristics. 

\subsection*{Measurements}

Measurements were performed in a dilution refrigerator with a lattice temperature of 25 mK. DC voltages were applied to the device using a multi-channel custom voltage source based on the Texas Instruments 1220 20-bit digital to analog converter. To form dots, the screening gates were grounded, a voltage of 2.8 V was applied to the accumulation gates (L/RA$_{1/2}$), and the plunger gates (P$_{1/2}$) were set to $\sim 2.1$ V forming a dot with approximately 30 electrons. For QD1, the depletion gates (L/RD$_1$) were set to $\sim$ 0.8 V to realize relatively low tunnel barriers and a device current on the order of 100 pA. In contrast, the tunnel barriers for QD2 were made larger by applying a voltage of $\sim$ 0.1 V, giving a tunnel rate $\sim$ 10 Hz for QD2 to the left reservoir and no tunnelling to the right reservoir, as described in the text. The plunger gates P$_{1/2}$ are then tuned on the mV scale to modulate the chemical potentials of the dots relative to the source and drain reservoirs. For QD1, we attempt to align the chemical potential to the reservoirs by maximizing the observed DC current at a finite bias. To measure the current hysteresis, a sinusoidal AC voltage with a frequency of 1 Hz is applied to the source of QD1 (L) and one of the depletion gates of QD2 (LD$_2$). AC and DC voltages are applied simultaneously to LD$_2$ using a custom-built voltage adder based on an AD797B ultralow-noise operational amplifier. The circuit bandwidth is $\sim$ 2 Hz due to RC filters on the lines connecting the external sources to the device, which limits the frequency of the applied voltage to $<2$ Hz. The 10 mV AC signal is applied directly to LD$_2$, but reduced by a factor of 10 using a voltage divider before being applied to L. The current through QD1 is amplified using a DL Instruments 1211 current-voltage preamplifier and then input to a Tektronix DPO 7104 oscilloscope. The oscilloscope recorded traces of length 10 s, with a sampling rate of 1 kS/s.

\subsection*{Master-equation model}

The Hamiltonian of the two-dot system reads
\begin{eqnarray}
H = \mu_\text{L} \ketbra{0,1}{0,1} + (\mu_\text{L} + E_\text{C}) \ketbra{1,1}{1,1},
\end{eqnarray}
where $\ket{n_1,n_2}$ denotes the state of the two-dot system, $n_1$ ($n_2$) is the occupation number of QD1 (QD2), and the energies of states $\ket{0,0}$ and $\ket{1,0}$ are zero. Following the approach in Ref.~\cite{Sun1999}, the evolution equation of the two-dot system reads
\begin{eqnarray}
\frac{d\rho_\text{S}}{dt} = -\frac{i}{\hbar} [H,\rho_\text{S}]
+ \mathcal{L}_1 \rho_\text{S} + \mathcal{L}_2 \rho_\text{S},
\end{eqnarray}
where
\begin{eqnarray}
\mathcal{L}_1 = \sum_{n_1,n_2 = 0,1} 2\gamma_1 [n_1 + (-1)^{n_1}f_1(n_2)]
\mathcal{D}[\sigma_{1-n_1,n_2;n_1,n_2}]
\end{eqnarray}
and
\begin{eqnarray}
\mathcal{L}_2 = \sum_{n_1,n_2 = 0,1} \gamma_2 [n_2 + (-1)^{n_2}f_2(n_1)]
\mathcal{D}[\sigma_{n_1,1-n_2;n_1,n_2}].
\end{eqnarray}
Here,
\begin{eqnarray}
f_1(n_2) &=& [ f(n_2E_\text{C}-\mu_\text{L})+f(n_2E_\text{C}-\mu_\text{R}) ]/2, \\
f_2(n_1) &=& f(\mu_\text{L}+n_1E_\text{C}), \\
\mathcal{D}[\sigma]\rho_\text{S} &=& \sigma \rho_\text{S} \sigma^\dag - \frac{1}{2} \{ \sigma^\dag\sigma , \rho_\text{S} \}
\end{eqnarray}
and
\begin{eqnarray}
\sigma_{n_1',n_2';n_1,n_2} = \ketbra{n_1',n_2'}{n_1,n_2}.
\end{eqnarray}

The rate equation for the occupation number of QD1 is
\begin{eqnarray}
\frac{d\bar{n_1}}{dt} = \abs{e}^{-1} ( I_\text{L} - I_\text{R} ),
\end{eqnarray}
where $I_\text{L}$ ($I_\text{R}$) is the current from reservoir-L to QD1 (from QD1 to reservoir-R). In the steady state, $d\rho_\text{S}/dt = 0$ and $I_\text{L} = I_\text{R}$. In the case that the voltage is switched with a frequency much lower than $\gamma_1$, the state of QD1 is approximately the steady state all the time, i.e.~$d\rho_\text{S}/dt \approx 0$, and the difference between $I_\text{L}$ and $I_\text{R}$ can be neglected. Currents are given by
\begin{eqnarray}
I_\text{L} = \sum_{n_1,n_2 = 0,1} \abs{e}\gamma_1 [n_1(-1)^{n_1} + f(n_2E_\text{C}-\mu_\text{L})]
\Tr(\sigma_{n_1,n_2;n_1,n_2}\rho_\text{S})
\end{eqnarray}
and
\begin{eqnarray}
I_\text{R} = - \sum_{n_1,n_2 = 0,1} \abs{e}\gamma_1 [n_1(-1)^{n_1} + f(n_2E_\text{C}-\mu_\text{R})]
\Tr(\sigma_{n_1,n_2;n_1,n_2}\rho_\text{S})
\end{eqnarray}
In our numerical simulations, we have taken $I = (I_\text{L} + I_\text{R})/2$.

\subsection*{Tunnel barrier resistance}

As described in the main text, a device imperfection led to a small tunnel barrier in series with QD1. To properly match the experimental results it is necessary to calculate the barrier resistance as a function of the voltage across it. The resistance is found using Ohm's law: $R_b(V_b) = V_b/I_b$, where $R_b$ is the resistance, $I_b$ is the current, and $V_b$ is the voltage. Current is calculated using the equation for transport between two thermally populated reservoirs \cite{Hu1987}:
\begin{equation}
I_b = \frac{\abs{e}}{h}\int_{-\infty}^{\infty}D(E)[f(E-V_b/2)-f(E+V_b/2)]dE
\label{eq:transmission}
\end{equation}
where $E$ is the energy, and $D(E)$ is the transmission coefficient. For simplicity, current is calculated for a square barrier, which has a transmission coefficient of:
\begin{equation}
D(E) = \exp(-\frac{4\pi{s}}{h}\sqrt{2m^*(\phi-E)})
\end{equation}
where $s$ is the width of the barrier, $m^*$ is the effective mass of the electron in Si, and $\phi$ is the height of the barrier in eV. The best match to the experimental data is found for $s = 40$ nm, and $\phi = 0.3$ meV.

\section*{Acknowledgements}

This work was supported by the EPSRC platform grant `Molecular Quantum Devices' (EP/J015067/1), and the EPSRC National Quantum Technology Hub in Networked Quantum Information Technology. The authors would like to thank Prof. Gerard Milburn for useful discussions. J.B. and G.W.H. would like to acknowledge the Quantum NanoFab facility at the University of Waterloo, and support from the Natural Sciences and Engineering Research Council of Canada, the Institute for Quantum Computing and the Waterloo Institute for Nanotechnology. 


\begin{thebibliography}{9}
\bibitem{footnote} The average current is calculated using $I_\text{ave}(t) = Q(t)/(N \delta t)$, where $Q(t)$ is the total charge going through the device in the time interval $[t,t+\delta t)$ in $N$ trials, where we have taken $N = 1000$ and $\delta t = 1\text{ ms}$.
\end{thebibliography}
\end{document}